\documentclass{article}

\usepackage{primarxiv}

\usepackage[utf8]{inputenc} 
\usepackage[T1]{fontenc}    
\usepackage{hyperref}       
\usepackage{url}            
\usepackage{booktabs}       
\usepackage{amsfonts}       
\usepackage{nicefrac}       
\usepackage{microtype}      
\usepackage{lipsum}
\usepackage{fancyhdr}       
\usepackage{graphicx}       
\usepackage{color}
\usepackage[table]{xcolor}
\usepackage{graphicx}
\usepackage{wrapfig}
\usepackage{caption}
\usepackage{subcaption}
\usepackage{adjustbox}
\usepackage{cuted}
\usepackage{amsmath} 
\usepackage{pbox}
\usepackage{stackengine}
\usepackage{ragged2e}
\usepackage{cite}

\definecolor{rev_color}{rgb}{0,0,0}

\newcommand\xrowht[2][0]{\addstackgap[.5\dimexpr#2\relax]{\vphantom{#1}}}

\title{Corticomorphic Hybrid CNN-SNN Architecture
for EEG-based Low-footprint Low-latency
Auditory Attention Detection}

\author{
  Richard Gall \\
  Department of Electrical and Computer Engineering \\
  University of Pittsburgh \\
  Pittsburgh, PA\\
  \texttt{rtg16@pitt.edu}\\
   \And
  Deniz Kocanaogullari \\
  Department of Electrical and Computer Engineering \\
  University of Pittsburgh \\
  Pittsburgh, PA\\
  \texttt{dek107@pitt.edu}\\
   \And
  Murat Akcakaya \\
  Department of Electrical and Computer Engineering \\
  University of Pittsburgh \\
  Pittsburgh, PA\\
  \texttt{akcakaya@pitt.edu}\\
   \And
  Deniz Erdogmus \\
  Department of Electrical and Computer Engineering \\
  Northeastern University \\
  Boston, MA, USA\\
  \texttt{erdogmus@ece.neu.edu}\\
   \And
  Rajkumar Kubendran \\
  Department of Electrical and Computer Engineering \\
  University of Pittsburgh \\
  Pittsburgh, PA\\
  \texttt{rajkumar.ece@pitt.edu}\\
}

\begin{document}
\maketitle

\begin{abstract}
In a multi-speaker "cocktail party" scenario, a listener can selectively attend to a speaker of interest. Studies into the human auditory attention network demonstrate cortical entrainment to speech envelopes resulting in highly correlated Electroencephalography (EEG) measurements. Current trends in EEG-based auditory attention detection (AAD) using artificial neural networks (ANN) are not practical for edge-computing platforms due to longer decision windows using several EEG channels, with high power consumption and large memory footprint requirements. Nor are ANNs capable of accurately modeling the brain’s top-down attention network since the cortical organization is complex and layered. In this paper, we propose a hybrid convolutional neural network-spiking neural network (CNN-SNN) corticomorphic architecture, inspired by the auditory cortex, which uses EEG data along with multi-speaker speech envelopes to successfully decode auditory attention with low latency down to 1 second, using only 8 EEG electrodes strategically placed close to the auditory cortex, at a significantly higher accuracy of 91.03\%, compared to the state-of-the-art. Simultaneously, when compared to a traditional CNN reference model, our model uses $\sim$15\% fewer parameters at a lower bit precision resulting in $\sim$57\% memory footprint reduction. The results show great promise for edge computing in brain-embedded devices, like smart hearing aids.\end{abstract}


\section{Introduction}

Approximately $35$ million Americans ($11.3\%$ of the population) suffer from hearing loss; this number is steadily increasing and is projected to reach $40$ million by 2025, according to a 2008 survey by Kochkin~\cite{kochkin2009marketrak}. This survey also revealed that only approximately $30\%$ of these individuals with hearing impairments prefer to use hearing aids, but hearing aid adoption is increasing with technological advances. For a healthy listener the ability to selectively attend to a particular speaker in a multiple-speaker or "cocktail party" \cite{cherry1953some} scenario is a trivial task. Those who suffer from hearing loss or are hearing impaired have difficulty with selective auditory attention \cite{shinn2008selective}. Cognition and selective auditory attention of an individual play important roles in listening and communication~\cite{shinn2008selective}, but unfortunately these are currently not fully utilized during hearing aid design~\cite{taylor2015does, kong2014classification}. 

State-of-the-art hearing aids are based on sophisticated digital signal processing (DSP) algorithms: (1) adaptive beamforming through directional microphones or binaural single microphones with wireless connection \cite{Dil01, Ric00, Ham02, Doe96, Kol93}; (2) adaptive noise reduction~\cite{Mar94, Ber79, Eph84, Mar02, Mar03n, Lot03}; (3) multiband compression and automatic gain control (AGC) for frequency specific and level dependent gain control~\cite{Sch97, Cor95, Byr01, Mar01, kong2012development, kong2013using}; and (4) context-aware AGC~\cite{Kat97, Pel02, Bre90, Wu03, kong2015effects}. Noise reduction distorts the spatial cues for directional listening~\cite{Bij10}. Using beamforming over monaural multiple microphones or binaural single or multiple microphones, hearing aids aim to achieve spatial selectivity~\cite{Dil01, Ric00, Ham02, Doe96, Kol93}. However, current methods lack the ability to reliably identify and amplify the sound source that is of interest to the hearing aid user. Accurate identification of the desired sound source (or its direction of arrival) is still an open research problem that needs to be solved for improved intelligibility of speech enhanced with hearing aids \cite{NIH}. Identification of the attended sound source becomes more challenging, especially during a dinner conversation, while driving, or in a noisy and crowded environment such as a cocktail party~\cite{Woo53, Ham04, Bij10, biesmans2015comparison, biesmans2017auditory, das2016effect}.

This makes a brain-embedded computer interface (BECI) hearing aid with the capability to assist in selective auditory attention desirable. In a multi-speaker "cocktail party" scenario, a listener is able to selectively attend to a speaker of interest, which is trivial for a healthy listener. This work aims to address selective auditory attention in listeners who suffer from hearing loss or are 
hearing impaired. Studies into the top-down auditory attention network of the brain have shown that it is possible to decode the auditory attention through the neural responses of a listener, suggesting that assisted selective auditory attention is possible \cite{yost1997cocktail, yost1993auditory, bronkhorst2015cocktail}. The top-down attention network consists of three areas: lateral intraparietal area, frontal eye fields, and the prefrontal cortex \cite{baluch2011mechanisms}. These areas of the brain encode spatial and temporal information with regard to the attention of the listener. The cortical responses of the listener were shown to be entrained to the speech envelope of the speaker of interest, resulting in neural activity that is correlated to the attended source's speech envelope \cite{horton2014envelope, o2015attentional, di2015low, ding2014robust}. The top-down attention modulation of the attended speaker's temporal envelope is the source of the correlation between neural activity and attended speaker's speech envelope \cite{ding2012emergence, kong2014differential, aiken2008human}. These effects have been observed in non-invasive electroencephalograph (EEG) and magnetoencephalography (MEG) \cite{lalor2010neural}. These observations have led to developments in auditory attention decoding (AAD) using EEG based BECI hearing aids \cite{fu2021auditory, haghighi2017graphical, cai2020low, an2021decoding}. The advances in AAD have shown that selective auditory attention of listeners can be inferred with relative accuracy from multi-channel EEG time-locked to auditory stimuli. EEG electrode based hearing aids are the preferred method of AAD as they are non-invasive and small in size.

Current trends in modeling the top-down attention network involve statistical or deep neural networks (DNN) models. However, neither the statistical nor the DNN models are practical for AAD hearing aids. The current results in studies involving statistical models rely on a large decision window typical in the range of 20-60 seconds \cite{haghighi2017graphical, haghighi2018eeg, o2014power, mirkovic2015decoding}. This is not practical for AAD hearing aids that have a desirable inference time of less than 5 seconds. DNN models are able to more accurately model the nonlinear nature of the auditory attention network. Due to this, the DNN models are able to have a smaller decision window that is in line with the design goals of the AAD hearing aids. However, these models have a undesirable low classification accuracy of around 80\% \cite{an2021decoding, cai2020low}. Furthermore, DNN models rely on a large number of parameters to better model the auditory attention network. The large number of parameters results in a large memory footprint, more compute resources (more MAC operations) and higher power consumption that is not realistic for BECI applications \cite{canziani2016analysis, xu2018scaling}. Additionally, both statistical and DNN models do not accurately model the top-down attention network. Statistical models do not address neither spatial nor temporal information of the listener and focus solely on the correlation of the speech envelopes and EEG signals. It is possible for DNN models to address both spatial and temporal encoding, however, conventional feed-forward ANNs with fully connected layers are not structurally compatible to the layered cortical processing happening in the brain.

One method for solving these problems is through the use of a deep spiking neural network (SNN). SNNs can deliver a scalable, hierarchical, sparse event-driven computing architecture inspired by retina and cortical structures in the brain for efficient local information processing on-the-edge. SNNs are suitable for edge computing with limited memory and power budget since the data pipeline has sparse activity and the weights can be stored with low-precision \cite{antelis2020spiking}. A SNN would utilize the same architecture as a standard ANN but replaces the activation functions with bio-realistic mechanisms, such as integrate-and-fire. A neuron will not activate, or fire, unless the inputs from the pre-synaptic neurons accumulate to greater than a threshold. This results in sparse, event-driven  computation that reduces the multiply-accumulation (MAC) operations and memory resources compared to that of an equivalent ANN. Furthermore a SNN with the same architecture as that of an ANN can be implemented, in hardware, with a significantly smaller memory footprint (due to the lower bit precision for the weights) and lower power consumption (due to the reduced number of MAC operations), while suffering from a reasonable loss in overall prediction accuracy \cite{qiao2015reconfigurable, essera2016convolutional, zhang2020neuro}. SNNs take advantage of sparse representations of the input signal along with low resolution weights, that introduce systematic (quantization) and random (spike jitter) noise into the system, in order to mimic dropout and stochasticity that has been proven to be useful for improving performance of standard ANNs. It achieves appreciable performance while being extremely energy-efficient and thus highly suitable for edge AI applications.

This paper aims to investigate whether it is possible to develop an AAD model capable of BECI applications, such as hearing aids, through the use of an energy efficient hybrid convolution neural network-spiking neural network (CNN-SNN) corticomorphic architecture. We define the term 'corticomorphic' to describe an AI model designed to more accurately mimic the layered cortical structures and signal processing in the brain. The proposed CNN-SNN architecture takes inspiration from the auditory cortex of the mammalian brain for top-down attention modulation  \cite{KELL2018630,drakopoulos2021convolutional}. The CNN-SNN model is capable of learning spatial-temporal features similar to those found in the auditory attention network of the brain, thereby providing improved performance in EEG-based AAD than current state-of-the-art approaches.

This paper is organized as follows. Section \ref{rel_work} presents an overview of related work for AAD using Statistical and Deep Neural Network (DNN) models. Section \ref{CorticoNN} describes the main contribution of this work, on Corticomorphic neural networks. Section \ref{methods} elucidates the methods used in this work to collect EEG data and process them using the proposed CNN-SNN model as well as a reference CNN model. Section \ref{results} presents detailed experimental results conducted with extensive hyperparameter tuning of the proposed models. Finally, Section \ref{summary} concludes the paper providing a summary of the achievements and shortcomings, leading to plans for future work.

\section{Related Work}
\label{rel_work}
In this section we discuss the two types of existing models for AAD: statistical and DNN models. Furthermore, the advances in modeling the auditory attention network are highlighted for each work.

\subsection{Statistical Models}

There are several studies into building statistical models for AAD. One study uses data from 128 EEG electrodes and dichotic stimuli (simultaneous stimulus to the right and left ear by different audio) \cite{o2014power}. Two different models are developed, one being subject-specific and the other a generalized model capable of working on any listener. This study shows that there is a higher cross-correlation between the attended speaker and the EEG signals captured. They are able to achieve an average accuracy of 89.0\% for the subject-specific model and 81.8\% for the generalized model while using a 60 second decision window.

Another study aims to reduce the number of EEG channels required for AAD as the previous result is not practical. The number of EEG channels are reduced to 96 and diotic stimuli (both sounds playing in each ear) replaces the dichotic stimuli from the last work \cite{mirkovic2015decoding}. Only subject-specific models are developed in this work. Using the decreased number of EEG channels, an average accuracy of 88.0\% across all subjects is achieved while keeping the decision window 60 seconds. Furthermore, additional experiments on the number of EEG channels used shows that increasing the number of channels to more than 25 has no significant affect on the accuracy of the model. Using as little as 5 channels, this work realizes an average accuracy of 74.1\%.

Further improving upon the results of the previous studies, another statistical model is developed aiming to reduce the number of EEG channels used and the decision window required. This work reduces the number of EEG channels used to 16 and they decision window to 20 seconds while still using diotic stimuli \cite{haghighi2017graphical}. Similar to the first, this study develops two types of AAD models: one being subject-specific and the other a generalized model. While reducing the number of EEG channels used and the decision window they show that generalized models are competitive with subject-specific models. The generalized model achieves an accuracy of 89.2\% and the subject-specific an average accuracy of 89.6\%. Though these results show a significantly high accuracy, subject-specific models are not practical for product design. 

\subsection{Deep Neural Network Models}

The previous experiments into developing statistical AAD models show the potential of AAD based hearing aids, but are not adequate due to the large decision windows. Therefore experiments developing auditory attention network models using DNNs are explored to better capture the complex spatiotemporal nature of the problem. One study uses a deep CNN architecture to attempt to better model the spatial nature of the auditory attention network \cite{an2021decoding}. This model uses three convolutional layers with average-pooling placed between them, with three fully-connected layers following. They develop a generalized model with a 1.2 second decision window using 64 EEG channels and diotic stimuli. The experiments show that the deep CNN AAD model is able to achieve an average accuracy of 77.0\% across all subjects. They reduce the inference time from the statistical models significantly but also reduce the accuracy.

A similar study attempts to further explore the spatial nature of the auditory attention network by introducing common spatial pattern (CSP) analysis of the EEG signals collected \cite{cai2020low}. This study also uses a deep CNN model, but reduces the number of parameters from the previous study. The architecture developed consists of a convolutional layer followed by an average-pooling layer and a fully-connected layer. They train two generalized models using data from 64 EEG channels and diotic stimuli with CSP applied and the other without CSP. Applying CSP on the EEG signals they achieve an accuracy of 82.1\% and 71.9\% accuracy for the model without, in a 5 second decision window. They further show the importance of spatial information in modeling the auditory attention network and reduced the number of parameters required compared to the previous study.

\section{Corticomorphic Neural Networks}
\label{CorticoNN}
Our work demonstrates a pioneering approach to developing hardware-friendly corticomorphic neural networks, to push the boundaries of ANNs. The architecture of our proposed hybrid network is inspired from the organization of the auditory and visual cortices found in the brain. From a neuroscience perspective, the divisions of the auditory cortex are the core (which includes primary auditory cortex, A1), the belt (secondary auditory cortex, A2), and the parabelt (tertiary auditory cortex, A3) \cite{pickles1998introduction}. The organization of neurons in the auditory cortex is according to the best response to the different frequency components of sound. At one end of the auditory cortex, the neurons respond best to low frequencies; while at the other end, neurons are sensitive to high frequencies. The multiple areas of the auditory cortex (similar to the multiple areas in the visual cortex), can be distinguished anatomically based on the observation of a complete "frequency map." This frequency map, also known as a tonotopic map, likely reflects the cochlear arrangement according to sound frequency. The primary tasks of the auditory cortex are in identification and segregation of "auditory objects" and sound source localization in space. For example, it has been shown that A1 encodes complex and abstract aspects of auditory stimuli without encoding their "raw" aspects like frequency content, presence of a distinct sound or its echoes \cite{chechik2012auditory}.

\begin{figure*} 
        \centering
        \vspace{-.05in}
        \includegraphics[width=\textwidth, keepaspectratio]{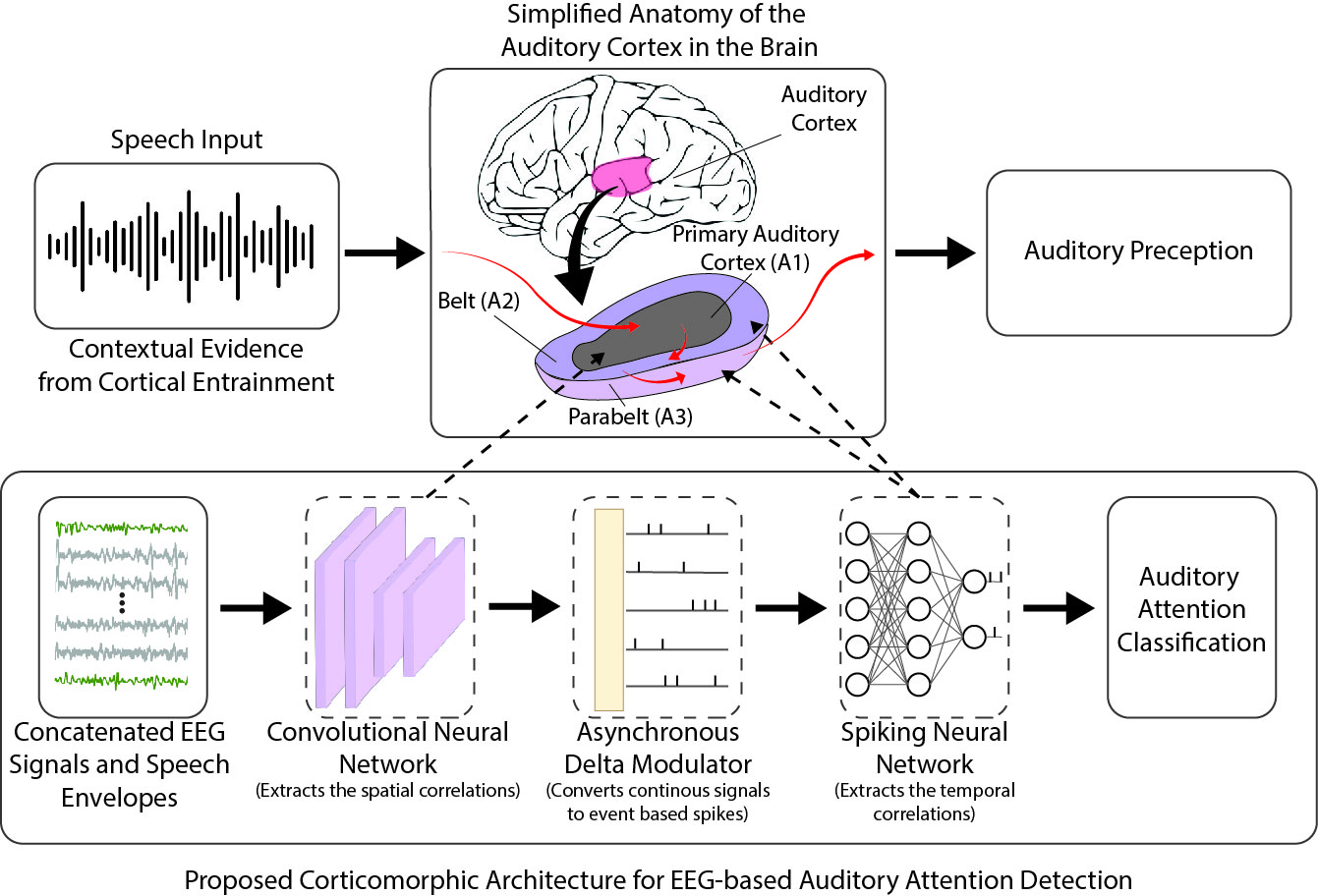}
        \caption{The proposed corticomorphic hybrid AI architecture, inspired by the anatomoy of the auditory cortex to implement EEG-based AAD. The pipeline for the auditory cortex is depicted by the red arrows. Speech input along with contextual evidence from other regions in the brain is passed into the primary auditory cortex (A1). The processed signals are then transmitted to the belt (A2), and finally the parabelt (A3) before it is perceived. Our hypothesis is that the convolution layer will mimic to the primary auditory cortex (A1) while the spiking neural network layer will emulate the belt (A2) and parabelt (A3) of the auditory cortex.}
        \label{fig:motivation}
\end{figure*}

The structure and function of the auditory cortex in humans has been extensively studied using functional magnetic resonance imaging (fMRI), EEG, and electrocorticography (ECoG). When trying to identify musical pitch, a peripheral part of this brain region is active, as indicated in human brain scans. As multiple sounds are transduced simultaneously in the hearing process, the role of the auditory system is to decide which components form the sound link. Since EEG recordings from the sites close to the auditory cortex can provide the peripheral activity of the region, our hypothesis is that there is high correlation between the EEG recordings and auditory attention to a particular speaker, since the EEG data provides contextual evidence to cortical entrainment to speech envelopes. Hence we use a CNN as the first stage in our hybrid model, which can extract correlation of input EEG waveforms along with speech envelopes of multiple speakers. We postulate that this CNN layer loosely emulates the primary cortex A1. 

Our model converts the output of the convolution layer to a train of spikes that can capture significant events in the output. By converting to spikes, we drastically increase the sparsity of information while only capturing what we need to process further, using a fully connected feed-forward SNN as the second stage of our proposed model. This reduces the memory and compute budget in the overall pipeline, while simultaneously improving the response time of the system by reducing the processing latency.

The proposed architecture and it's relation to the auditory processing of the human brain is depicted in Figure \ref{fig:motivation}. This figure highlights the cortical pathway that the brain takes when processing auditory stimulus (depicted in red arrows) and how the corticomorphic architecture is structured to model this processing. When processing auditory input, the stimulus is passed in to the primary auditory cortex (A1), the belt (A2), and the parabelt (A3), respectively. We postulate that using the CNN-SNN corticomorphic architecture, the proccesing pipeline of the brain will be accurately modeled. The CNN layer will be capable of modeling the primary auditory cortex (A1), where the spatial information is extracted from the stimulus. The belt (A2) and the parabelt (A3) are responsible for the extraction of temporal information, the SNN layer will be able to accomplish this through the spike encoding. The resulting architecture models the auditory cortex making this model efficient in AAD. 

The proposed model aims to improve upon the results of the past studies which used Statistical and DNN based models. The corticomorphic CNN-SNN model architecture will be more biologically accurate to the auditory attention network. Through the use this architecture we will investigate the ability to reduce the number of EEG electrodes used and the decision window compared to those seen in the previous works. The SNN layers will explore the temporal relationship of AAD whereas the previous DNN models only consider the spatial relationship. Additionally the CNN-SNN model can be practically realized on BECI (such as smart hearing aid design for AAD) due to the low-power SNN hardware implementation, which the previous studies are not capable of.

\section{Methods}
\label{methods}
\subsection{EEG Collection and Processing}

Ten volunteers (5 male, 5 female), between the ages of 25 to 30 years, with no known history of hearing impairment or neurological problems participated in this study, which followed an IRB-approved protocol. EEG signals were recorded using a g.USBamp biosignal amplifier using active g.Butterfly electrodes with cap application from g.Tec (Graz, Austria) at 256 Hz. Sixteen EEG channels (P1, PZ, P2, CP1, CPZ, CP2, CZ, C3, C4, T7, T8, FC3, FC4, F3, F4 and FZ according to International 10/20 System) were selected to capture auditory function related brain activities over the scalp. Signals were filtered by built-in analog bandpass ([0.5, 60] Hz) and notch (60Hz) filters to remove frequency components that are not relevant.  aThe signals were filtered further, after acquisition and noise removal, with a linear-phase 128 order FIR filter ([.5, 40] Hz) using the DSP System Toolbox from MATLAB. The speech envelopes were calculated using the Hilbert transform of the auditory stimuli.

\begin{figure}
     \centering
    \includegraphics[width=\columnwidth, keepaspectratio]{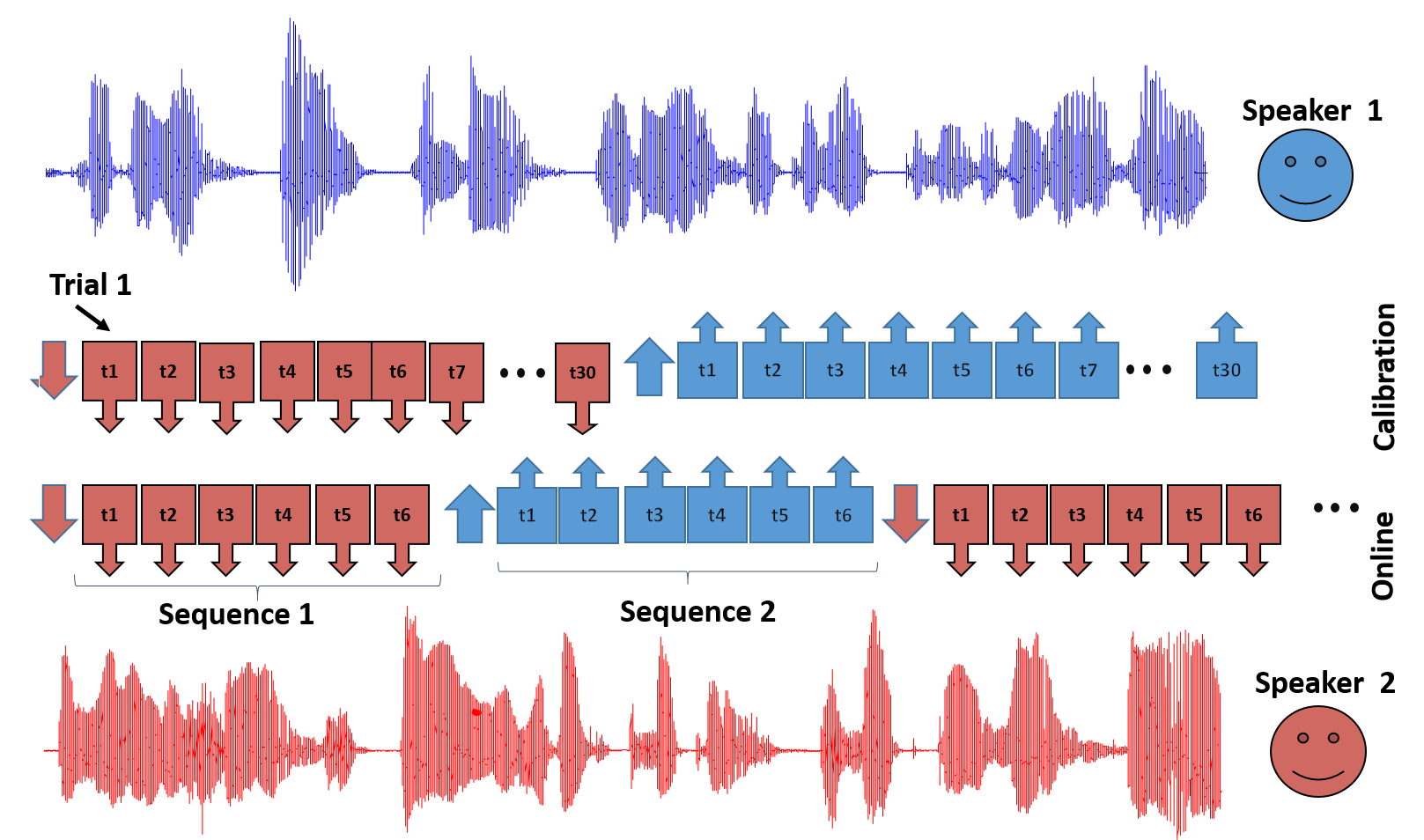}
    \caption{The experimental paradigm of the calibration and online sessions visualized \cite{haghighi2017graphical}. The top row represents the calibration session and the bottom row represents the online session. Each box represents a trial of 20 seconds with the arrow direction indicating the speaker of interest.}
    \label{fig:expParad}
\end{figure}

Each participant attended a 30 minute calibration session. Each session consisted of 60 trials of 20 seconds of diotic auditory stimuli with 4 seconds breaks between each trial. The diotic auditory stimuli are generated by one male and one female speaker. These speakers narrated a story (different story for different speakers chosen from audio books of literary novels) for 20 minutes. We consider every 20 seconds of this 20-minute-long diotic narration as a trial. During this session, participants were asked to passively listen to 20-minute-long narration, and they were instructed to switch their attention from one speaker to another during different trials.  The instructions to switch attention from trial to trial are provided to the user on a computer screen using "f" and "m" symbols for female and male speakers, respectively. Following the calibration session each subject participated in an online session. The online session consisted of 20 sequences that were 2 minutes long and held 6 trials each, 20 seconds in length. The speech waveform energies were normalized to ensure each trial had equal energy for both speakers. Between trials, a target speaker was indicated by displaying a "f" or "m" on the screen to indicate whether to attend to the female or male speaker. The experimental paradigm is summarized in Figure \ref{fig:expParad}, where the first row visualizes the calibration session and the second row illustrates the online session. The direction of the arrows indicate the attended speaker for each trial. This data set has been used successfully in several previous works \cite{haghighi2017graphical, haghighi2018eeg}.

\subsection{CNN-SNN}

\begin{figure} 
        \centering
        \includegraphics[width=0.5\columnwidth, keepaspectratio]{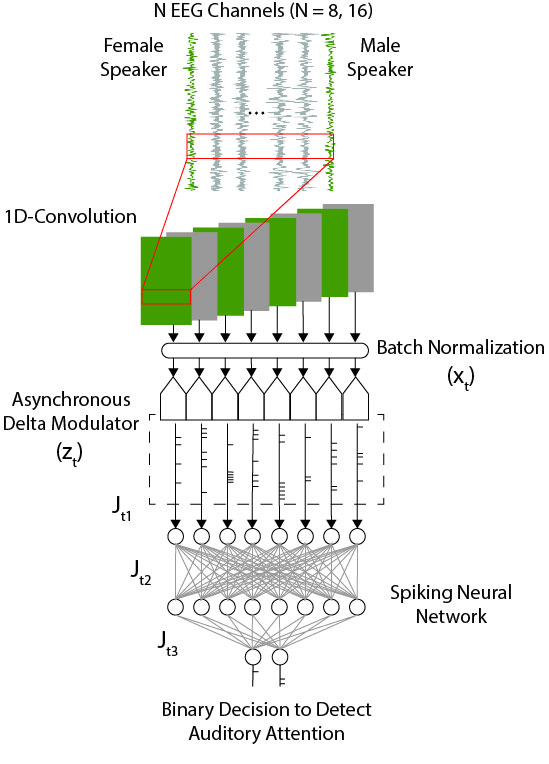}
        \caption{The proposed CNN-SNN architecture illustrated using a top to bottom signal flow diagram. Waveforms on the top represent the input to this model, where the green traces are the speech envelopes and the gray traces are the collected EEG signals. This concatenated signal is passed into the 1D-convolution layer to capture the spatial correlation between the EEG signals and the speech envelope of interest. The output of the convolution layer is normalized using a batch normalization layer before being converted to ON and OFF events using the ADM. These spikes are then processed by the two fully-connected SNN layers to get the final classification output.}
        \label{fig:CNNSNNModel}
\end{figure}

The CNN-SNN architecture illustrated in Figure~\ref{fig:CNNSNNModel} was developed to better model the spatiotemporal cortical processing of top-down attention network. The first stage of the proposed architecture consists of a Convolution layer in order to capture spatial information encoded in the EEG signals. Following the convolution layer is a SNN layer that forms the second stage. Adding the SNN layer helps in modeling the temporal spike encoding of the top-down attention network more accurately. The corticomorphic architecture comprising these two stages should be capable of mimicking the bio-realistic spatiotemporal features found in the auditory attention network found in the mammalian brain. The model consists of a 1D-convolution layer, a batch normalization layer, and two fully-connected spiking layers (Input:80, Hidden: 80, Output:2). The convolution layer has a kernel size of 64, representing a 0.25s window, and 40 output channels. This layer does not include an activation function in order to fully capture the negative and positive spatial correlations from the input. The SNN layers use leaky-integrate and fire (LIF) neurons with dynamics governed by Equation 1, where $\tau_{RC}$ is the membrane time constant, $V_t$ is the neuron's membrane voltage at time t, $J_{tl}$ is the accumulated input current from the \emph{l}  layer at time t and $\delta _t$ is the change in time from the previous time step \cite{hunsberger2016training}. $\tau_{rc}$ is assumed to be fixed to 20 ms, however, a grid search can be performed to optimize this hyper-parameter. 

\begin{align}
    v_{t} = v_{t-1} - (J_{tl} - v_{t-1})e^\frac{\delta _{t}}{\tau _{RC}}\\
    if v_{t} > 1 V, \text{ a spike is registered}\nonumber
\end{align}

\begin{figure}[h]
\vspace{-2em}
     \centering
    \includegraphics[width=0.5\textwidth]{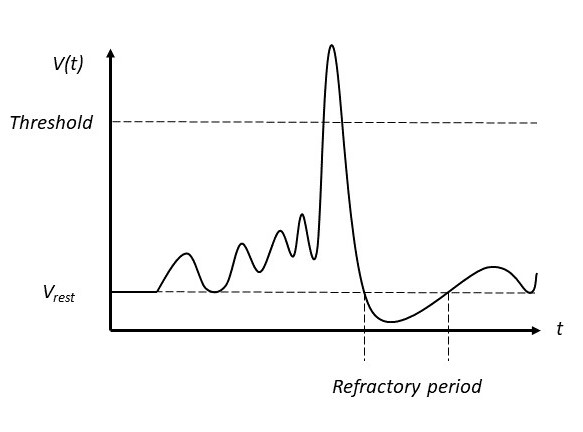}
    \caption{An illustration of a spiking neuron's operating mechanism. The accumulation of spike inputs from the pre-synaptic neurons causes the voltage, V(t), to change until it reaches a threshold that registers an output spike. Following the spike the neuron enters the refractory period where it cannot spike again before resetting to the resting potential, $V_{rest}$, and allowing the accumulation process to start again.}
    \label{fig:neuron}
\end{figure}

\noindent If the LIF neuron accumulates (or integrates) sufficient charge from the pre-synaptic spike activity such that the membrane voltage crosses a threshold more than 1 V, it will 'fire' an action potential or 'spike'. After firing a spike, the membrane potential of the neuron will reset to 0 V. The neuron will not be able to spike again until the refractory period has passed, regardless of the inputs from the pre-synaptic neurons. The mechanism of the aforementioned neuron are illustrated in Figure \ref{fig:neuron}. In this figure, the accumulation of the pre-synaptic inputs are shown as slight increases in the voltage over time, with leakage trying to decrease the potential. When there is sufficient input spike activity called 'events', the membrane potential will surpass the threshold. After surpassing the threshold of the neuron, 1 V in our case, a spike will be registered and this spike will be passed to the connecting neurons in the downstream pipeline. Following the spike, the neuron enters it's refractory period during which it is unable to fire another spike. 
After the refractory  period, the neuron then resets to it's resting potential, which is 0 V for the LIF neuron used in this simulation model, before starting the accumulation or integration process again.

In order to train this hybrid model, a CNN model was initially trained independently. This model consisted of a 1D-convolution layer, a batch normalization layer, an average pooling layer, and two fully connected layers. This model uses the same configuration as the hybrid CNN-SNN model described above except that an average-pooling layer is added after the convolution layer and a standard fully connected layer replaces the spiking neural network. After successful training resulting in convergence (i.e. loss is minimized and accuracy is maximized), any layers not present in the CNN-SNN model are removed. This modified network is used for inference, where the EEG data with the two speakers' speech envelopes concatenated to both sides is fed into the first stage comprising the convolution layer. The output of the first stage is converted into spike events using an asynchronous delta modulator (ADM). The ADM compares the values of the first stage output, during the previous time step to the current time step and generates ON/OFF events based on large changes in the output value between the time steps. The values of the output of the first stage, before the EEG signal is presented, are assumed to be zero. An ON event represents a positive change i.e. increase in output value greater than a chosen threshold, while an OFF event represent a negative change i.e. decrease in output value greater than the negative of the same threshold. If the change from one time step to the next does not meet these requirements, meaning the output value has not increased or decreased sufficiently, then no event is registered. 

Assigning a low threshold generates too many spike events that capture more information than necessary, resulting in increased computation and hence higher power consumption. On the other hand, if a high threshold is used, the ADM generates very few spikes with significant loss in information, resulting in poor classification accuracy. Therefore, a grid search was used to find the optimal threshold for the ADM to encode the spikes sparsely while still carrying enough information to provide maximum accuracy while consuming minimal compute resources and hence lower power consumption. By adding the batch normalization layer, the outputs of the CNN layer will be normalized to make the grid search easier and more consistent. The result of the grid search depends on the weight initialization for a training session, however, it was found that between 0.5 and 0.4 is the ideal threshold for the ADM. The sparse events generated by the ADM were used to train the SNN independently using Nengo's deep learning library \cite{Rasmussen2018}.  Equations 2, 3, and 4 summarize the ADM's operations, where $x_{t}$ is the ADM input provided by the CNN layer at time t, T is the threshold as determined by the grid search, and $z_{t_{ON/OFF}}$ is the output of the ADM with the respective event type at time t. The $z_{t_{ON}}$ and $z_{t_{OFF}}$ events are concatenated and used as the input to the first dense spiking layer of the SNN. The concatenated inputs of $z_{t_{ON/OFF}}$ for a given neuron will be accumulated and the resulting summation is denoted by $J_t$, for each neuron.

\begin{align}
    \delta _{x} = x_{t} - x_{t-1}\\
    \delta _{x} > T, z_{t_{ON}} = 1\\
    \delta _{x} < -T, z_{t_{OFF}} = 1
\end{align}

\noindent The final time step from the decision window is used as the final classification for the attended speaker. The Nengo deep learning library approaches training of the non-differentiable neurons through surrogate-models that are differentiable \cite{hunsberger2016training}. The non-differentiable neurons are replaced during training with an approximate model that is differentiable. This allows for backpropagation to still apply for SNN training and the surrogate-models are replaced, after training, with the LIF neuron model to generate discrete spiking impulse train, based on the Dirac delta function, $\delta(t)$. Training of the SNN was done using ADAM optimizer and cross-entropy loss function. 

\subsection{CNN}

A traditional CNN model was developed as a control reference for the performance of the proposed hybrid architecture. The model's dimensions were selected to keep the number of parameters similar to that of the CNN-SNN model. However, due to the ADM generating ON and OFF events for each output channel of the convolution layer, the inputs to the fully connected layer in the CNN-SNN model is twice that of the standard CNN model used here. This model is comprised of a 1D-convolution layer, average-pooling, and two fully-connected layers (Input:40, Hidden: 40, Output:2). The convolution layer is the same as described in the previous sub-section, with a kernel size of 64, representing a 0.25s window, and 40 output channels. This layer does not include activation function to keep the two models consistent. The average-pooling layer kernel size was selected to keep the input size to the first fully-connected layer consistent among different decision windows. Rectified Linear Unit (ReLU) activation function was used for the first fully-connected layer and Softmax for the second. Cross-entropy was the loss function used for training with ADAM as the optimizer. The input to this model is the EEG data collected with the two speakers' envelopes concatenated on either end, which is the same as the CNN-SNN model.

This model is trained in two scenarios, one with Least Absolute Shrinkage and Selection Operator (LASSO) regression applied on the activation outputs and one without LASSO regression. LASSO was used to regularize the model and sparsify the weights. The LASSO or L1 penalty is applied as follows,

\begin{align}
    Loss &= E(y, \hat{y}) + \lambda_{L1}\sum |\theta |
\end{align}

\noindent E(y, $\hat{y}$) is the error between the desired output y and the predicted output $\hat{y}$, as calculated by the cross-entropy loss function. The LASSO penalty, $\lambda_{L1}$, scales the loss by the sum of the weights, $\theta$. $\lambda_{L1}$ is selected to be the largest possible until the model is no longer able to learn due to the scaling of the loss. Trials are conducted for each decision window in order to find the optimal $\lambda_{L1}$ to achieve this target.

\section{Experimental Results and Discussion}
\label{results}
\begin{figure*}
    \begin{subfigure}[c]{.5\textwidth}
        \includegraphics[width=\linewidth, keepaspectratio]{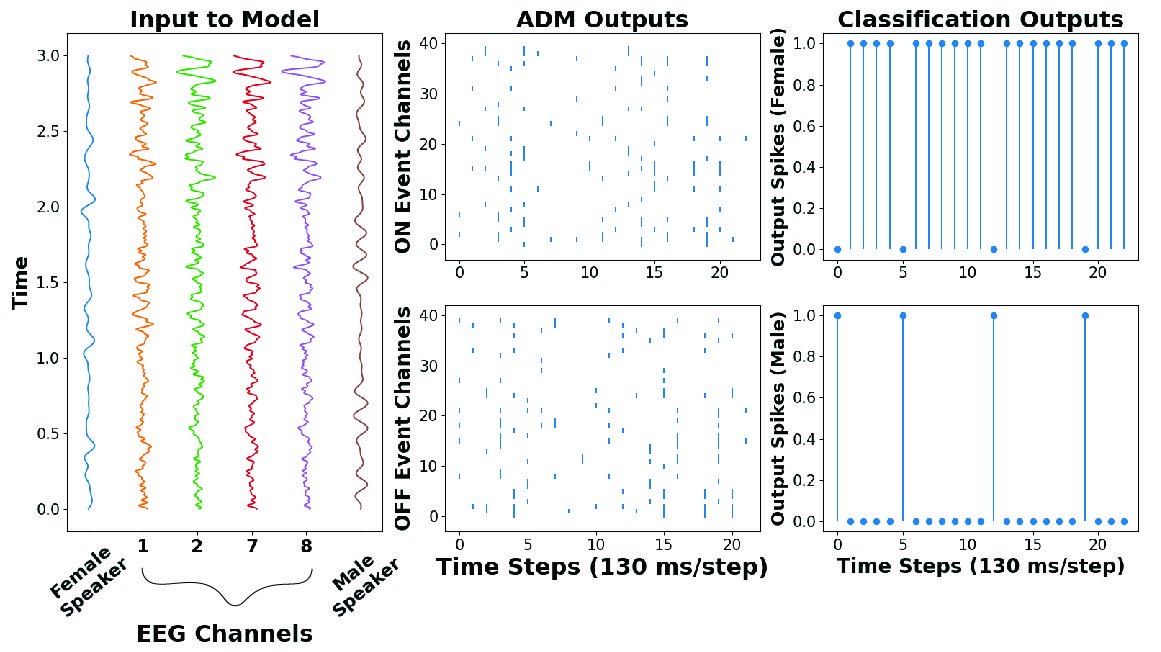}
        \caption{Subject 1 attending to the female speaker.}
        \label{fig:femalespeaker}
    \end{subfigure}
    \begin{subfigure}[c]{.5\textwidth}
        \includegraphics[width=\linewidth, keepaspectratio]{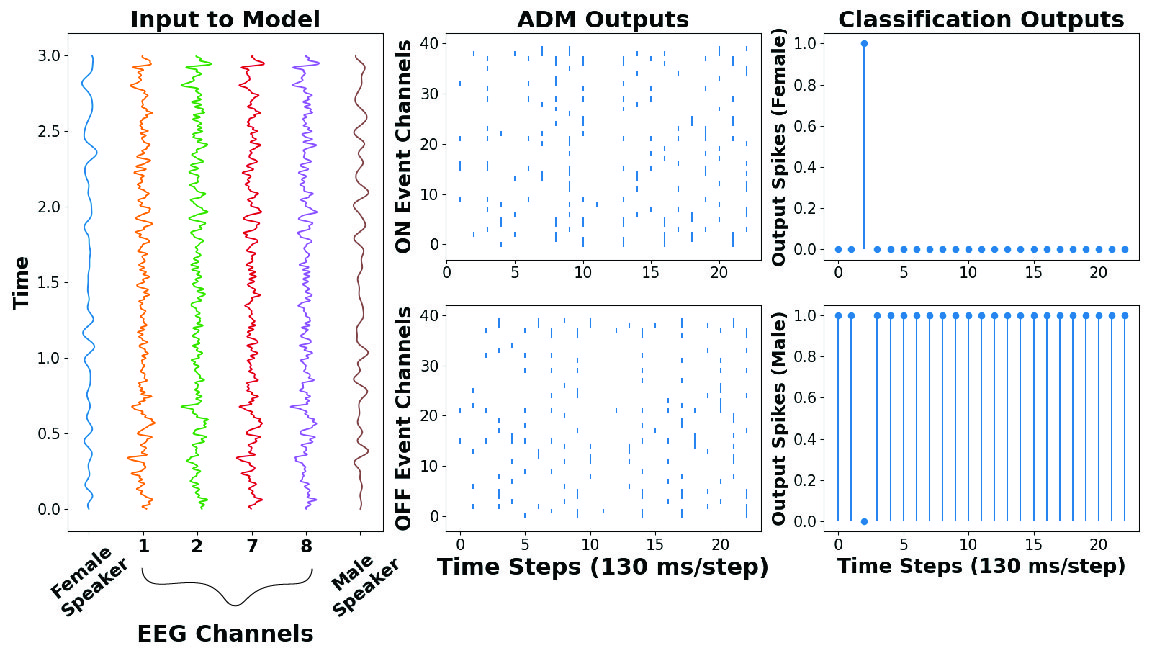}
        \caption{Subject 1 attending to the male speaker.}
        \label{fig:malespeaker}
    \end{subfigure}
    \begin{subfigure}[c]{.5\textwidth}
        \includegraphics[width=\linewidth, keepaspectratio]{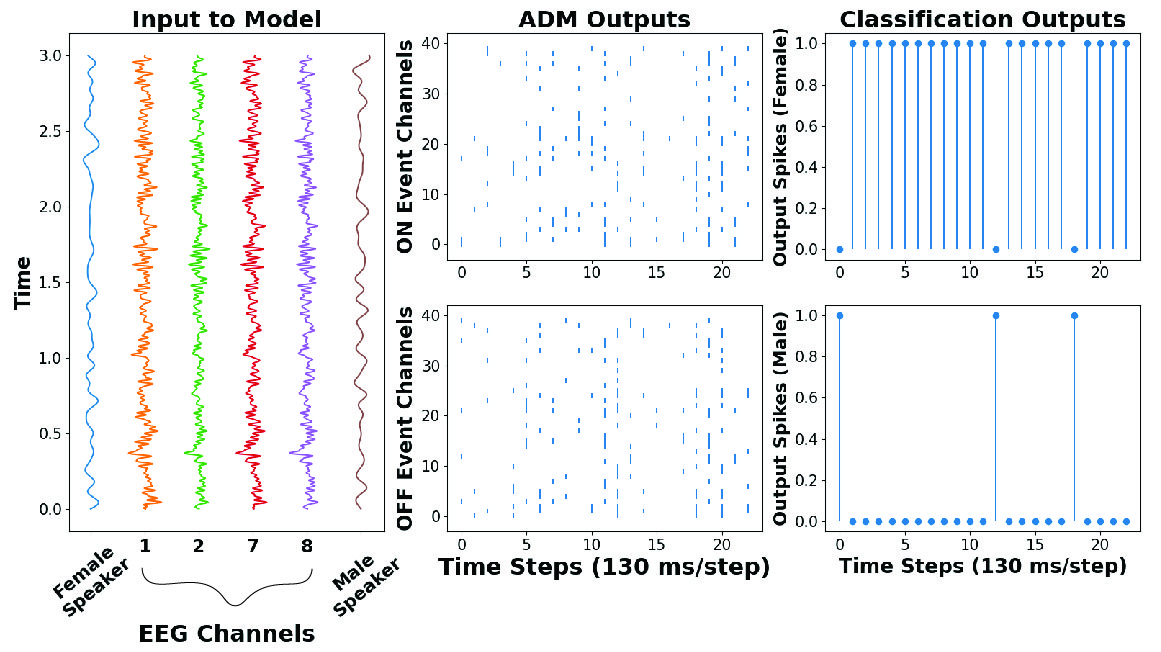}
        \caption{Subject 2 attending to the female speaker.}
        \label{fig:femalespeaker2}
    \end{subfigure}
    \begin{subfigure}[c]{.5\textwidth}
        \includegraphics[width=\linewidth, keepaspectratio]{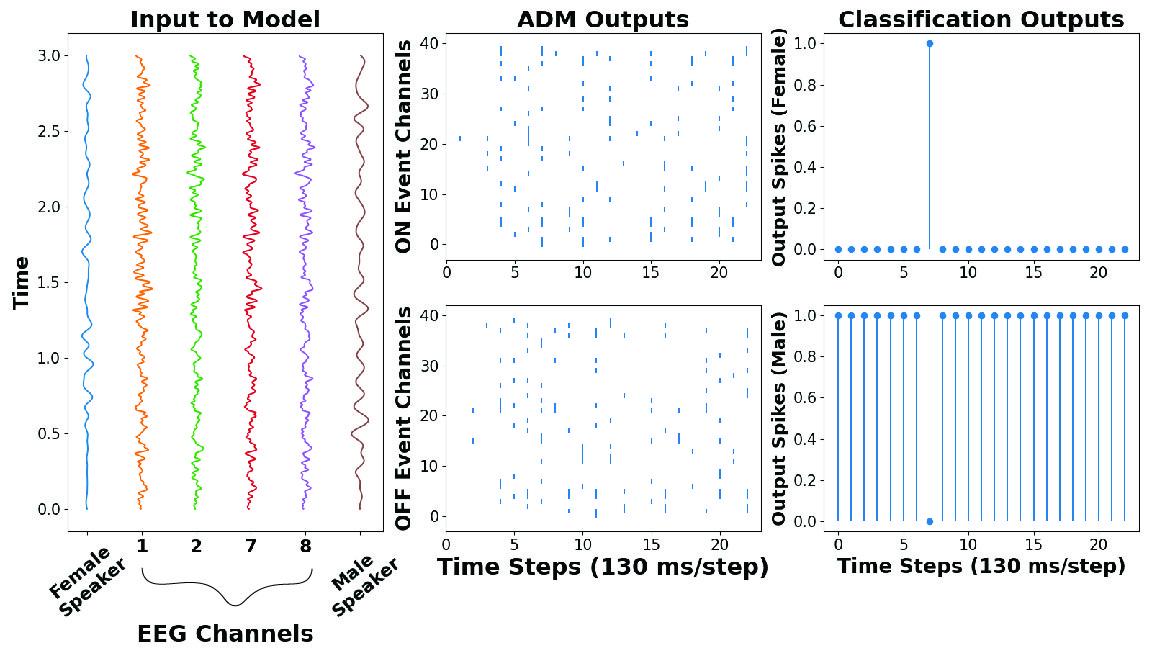}
        \caption{Subject 2 attending to the male speaker.}
        \label{fig:malespeaker2}
    \end{subfigure}
    \caption{Experimental results for two different subjects attending to the female (a)(c) and male (b)(d) speaker for a 3 second decision window, shown as four sub-figures. In each sub-figure, the order of the stages flow from left to right. The left most figure represents the concatenated EEG signal and speech envelope input to the model with only selected EEG channels shown. The output of the ADM stage with 40 channels corresponding to the output of the convolution layer are shown in the middle figure, where the top raster plot depicts the ON events and the bottom raster plot the OFF events. The right most figure presents the final classification output spikes for each time step. The top and bottom inset plots represent the two possible speakers: female or male, where the classification at the final time step is used as the result from the model.}
    \label{fig:modelStages}
\end{figure*}

The two networks are trained with several variations, using 16 EEG channels and using 8 EEG channels for several different decision windows times. The 8 EEG channels are selected as a subset from the original 16 (P1, PZ, P2, CP1, CPZ, CP2, CZ, C3, C4, T7, T8, FC3, FC4, F3, F4 and FZ by the International 10/20 System) based on the proximity to the auditory cortex. The 8 selected from the original 16 channels were C3, C4, CZ, CPZ, CP1, CP2, P1, and, P2. For both CNN and CNN-SNN models, training was done using ADAM optimizer and cross-entropy loss function. A train-validation split of 80\%-20\% is performed on the calibration data set for training the models. During training phase, the model is run against the validation set for each epoch. The model with the best validation accuracy is selected as the final model to be used in the testing phase. The testing data set, which is entirely different from the training data, consists of the full online session data to ensure a large number of samples could be used for model evaluation. 


The different stages of the CNN-SNN model are depicted in Figure \ref{fig:modelStages}, for the case of subject 1 and 2 attending to the female speaker and male speaker for a 3 second decision window. The first stage shows the concatenated speech envelopes and EEG signals of selected few channels for ease of visualization when plotted. Following the first stage is the Asynchronous Delta Modulator (ADM) stage, converting the CNN output into ON and OFF events. It is possible to see the sparse encoding from the ADM stage by comparing the classification outputs (output of second stage, SNN) between Figure \ref{fig:femalespeaker}\textbackslash\ref{fig:femalespeaker2} and Figure \ref{fig:malespeaker}\textbackslash\ref{fig:malespeaker2}. The sparse encoding seen in these figures is a consequence of the grid search performed to ensure the ADM outputs a unique pattern for each time step. This unique pattern allows the model to be trained to accurately decode the attention of the listener. The classification output with two channels, one for the male and one for the female speaker, is the final stage of the architecture. The spike outputs represent which speaker the model classified as the attended speaker at each time step. From Figure \ref{fig:modelStages}, it can be readily inferred that the proposed corticomorphic hybrid CNN-SNN correctly identified the attended speaker at the last time step. Furthermore, the model accurately identified the correct speaker for almost all time steps when paying attention to the male speaker or the female speaker. One reason for the instances of misclassifications could be due to the threshold of the ADM being too large to optimally encode the critical information when converting into spikes. As illustrated in these examples, a change from an ON/OFF/NO event to another event could result in a change from misclassification to a correct classification. This can be achieved by tightening the grid search so that the ADM could be better optimized to encode the data. Tightening the grid search is not required, however, since these classification errors are occurring in time steps prior to the final classification output at the end of the decision time window. Additionally, due to the ADM assuming the values prior to the EEG data as zero, the first time step is not encoded with respect to the attention of the listener. Due to the missing information from the ADM the model classifies as male for the first time step in each example. This is corrected after the first time step as the ADM is now able to use actual values from the CNN which processed the valid EEG data, rather than the initially assumed zero.

The performance results of the proposed hybrid CNN-SNN and reference CNN models are shown in Table \ref{tab:8channels} and Table \ref{tab:16channels} for 8 and 16 EEG channels, averaged over 20 random training initializations. Comparing the results between the 8 and 16 EEG channel tables it is clear that using 8 EEG channels has a similar accuracy to the 16 EEG channels for Auditory Attention Detection (AAD). This suggests that having a few well-placed electrodes, centered around the auditory cortex can achieve a comparable accuracy to having a large number of EEG channels. Our observation is further validated because adding additional EEG channels introduces unnecessary information that is not needed for decoding the attention of a listener. Having fewer EEG channels is an important factor for the design goal of an efficient hearing aids for EEG-based AAD as ultimately the hearing aid can only house a few electrodes. The 8 EEG channel models also require less number of parameters for the convolution layer due to the smaller input size of the concatenated EEG signals and speech envelopes. This is crucial as it can significantly reduce the memory footprint, computational resources and overall power consumption of the model when ported onto hardware, another important factor of energy-efficient AAD in hearing aids. Furthermore, both the 8 EEG channel and 16 EEG channel results demonstrate increasing accuracy as the decision window decreases. The additional information from a larger decision window is the probable causes for this observation. Since the models used here are primarily feed-forward networks (convolution and fully connected layers in CNN and SNN) with no recurrent connections, they have no memory from previous time steps. Hence it is likely that the larger decision window with a higher number of time steps introduces too much information for the model to decode, making it harder to find the global minimum during training. It is preferable to have a smaller decision window for hearing aids as the detection of attention can be updated frequently to enable real-time adaptation of attention to different speakers. Therefore, the low latency in detecting auditory attention is the primary advantage enabled by our proposed model. Hence, the proposed corticomorphic hybrid CNN-SNN architecture achieves efficient AAD with low footprint (memory and compute) and low latency, which is not possible with other state-of-the-art implementations.


\begin{table}
    \begin{center}
        \huge
        \begin{adjustbox}{width=0.5\columnwidth, center}\begin{tabular}{|c|c|c|c|c|c|c|}
            \multicolumn{7}{c}{\textbf{8 EEG Channels}} \\
             \hline\xrowht{30pt}
              & \multicolumn{2}{c|}{CNN} & \multicolumn{2}{c|}{CNN, LASSO} & \multicolumn{2}{c|}{CNN-SNN} \\
             \hline
             \pbox{15cm}{Window\\ Length (s)} & Accuracy & F1-Score & Accuracy & F1-Score & Accuracy & F1-Score \\
             \hline\xrowht{30pt}
             5 & 72.90\% & 74.20\% & 70.90\% & 70.15\% & 83.56\% & 82.83\% \\
             \hline\xrowht{30pt}
             4 & 73.71\% & 75.44\% & 71.55\% & 71.78\% & 84.29\% & 84.98\% \\
             \hline\xrowht{30pt}
             3 & 74.64\% & 72.88\% & 73.50\% & 72.59\% & 87.91\% & 87.73\% \\
             \hline\xrowht{30pt}
             2 & 78.77\% & 77.86\% & 75.40\% & 76.27\% & 89.78\% & 89.85\% \\
             \hline\xrowht{30pt}
             1 & 79.01\% & 79.04\% & 77.69\% & 78.49\% & 91.03\% & 91.07\% \\
             \hline
        \end{tabular}
        \end{adjustbox}
        \vspace{2mm}
        \caption{Comparison of the accuracy and F1 scores for the CNN, CNN with LASSO, and CNN-SNN models using 8 EEG channels, averaged across 20 random initialization of training, testing and validation runs.}
        \label{tab:8channels}
    \end{center}
\end{table}


\begin{table}
    \begin{center}
        \huge
        \begin{adjustbox}{width=0.5\columnwidth, center}\begin{tabular}{|c|c|c|c|c|c|c|}
            \multicolumn{7}{c}{\textbf{16 EEG Channels}} \\
             \hline\xrowht{30pt}
              & \multicolumn{2}{c|}{CNN} & \multicolumn{2}{c|}{CNN, LASSO} & \multicolumn{2}{c|}{CNN-SNN} \\
             \hline
             \pbox{12cm}{Window\\ Length (s)} & Accuracy & F1-Score & Accuracy & F1-Score & Accuracy & F1-Score \\[.5ex]
             \hline\xrowht{30pt}
             5 & 71.74\% & 72.96\% & 69.86\% & 67.92\% & 82.09\% & 83.41\%  \\
             \hline\xrowht{30pt}
             4 & 73.59\% & 74.78\% & 71.58\% & 73.13\% & 83.11\% & 83.54\% \\
             \hline\xrowht{30pt}
             3 & 74.00\% & 72.80\% & 72.48\% & 71.14\% & 85.27\% & 85.00\% \\
             \hline\xrowht{30pt}
             2 & 75.85\% & 74.47\% & 73.43\% & 72.88\% & 88.37\% & 88.18\% \\
             \hline\xrowht{30pt}
             1 & 76.48\% & 77.07\% & 74.75\% & 75.29\% & 89.35\% & 89.57\% \\
             \hline
        \end{tabular}
        \end{adjustbox}
        \vspace{2mm}
        \caption{Comparison of the accuracy and F1 scores for the CNN, CNN with LASSO, and CNN-SNN models using 16 EEG channels, averaged across 20 random initialization of training, testing and validation runs.}
        \label{tab:16channels}
    \end{center}
\end{table}


\begin{table}[!h]
    \centering
    \begin{adjustbox}{width=0.45\columnwidth, center}\begin{tabular}{|c|c|c|c|}
        \hline\xrowht{15pt}
         Model & \pbox{10cm}{Number of\\ Electrodes} & \pbox{10cm}{Decision\\ Window (s)} & Accuracy \\
         \hline\xrowht{8pt}
         Statistical \cite{o2014power} & 128 & 60 & 89.0\% \\
         \hline\xrowht{8pt}
         Statistical \cite{mirkovic2015decoding} & 96 & 60 & 88.0\% \\
         \hline\xrowht{8pt}
         Statistical \cite{haghighi2017graphical} & 16 & 20 & 89.6\% \\
         \hline\xrowht{8pt}
         CNN \cite{an2021decoding} & 64 & 1.2 & 77.0\% \\
         \hline\xrowht{8pt}
         CNN+CSP \cite{cai2020low} & 64 & 5 & 82.1\% \\
         \rowcolor{lightgray}\hline\xrowht{8pt}
        CNN-SNN & 8 & 1 & 91.03\% \\
         \hline
    \end{tabular}
    \end{adjustbox}
    \vspace{2mm}
    \caption{Comparison of the accuracy and F1 scores of the best results from the related worked section with the number of EEG channels and decision window used. The best result from this work is highlighted in gray.}
    \label{tab:comparAcc}
\end{table}

By modeling the auditory attention network of the brain with a similar topology, the corticomorphic architecture outperforms the CNN model developed for this study. This is a promising result for designing AAD in hearing aids as this architecture is capable of low power computation and small memory footprint making it ideal for BECI applications. The CNN-SNN model was able to achieve a 10\% or greater accuracy for each decision window compared to the control reference CNN model, which is a significant accuracy increase. The CNN-SNN model does require around an additional five thousand parameters as a consequence of having ON and OFF events generated out of the ADM and the additional batch normalization layer for optimizing the grid search. However, it is possible to overcome this shortcoming by reducing the network and the bit precision of the weights in order to use fewer parameters and lesser memory to store the weights. 



A comparison between the results of the proposed corticomorphic CNN-SNN architecture and the results of the previous works are summarized in Table \ref{tab:comparAcc}. This table highlights three important factors of each study: the number of EEG electrodes used, the length of the decision window, and the accuracy of the models developed. Compared to the previous works in AAD, this study used the smallest number of EEG electrodes for decoding the attention of a listener. All of the previous works, other than a statistical model \cite{haghighi2017graphical}, used 64 or more electrodes; whereas this study used only 8 electrodes for demonstrating the highest accuracy. Moreover, the CNN-SNN is the only model with a practical decision window size of 1 second for AAD in hearing aids, aside from the CNN model developed in \cite{an2021decoding}. The CNN-SNN model not only used less EEG channels and a smaller decision window, it also improved the state-of-the-art accuracy. This study achieved an accuracy of 91.03\%, compared to the state-of-the-art statistical model \cite{haghighi2017graphical} our architecture achieves a 5\% increase in accuracy while using a 20x smaller decision window and 2x less EEG electrodes. Additionally, the state-of-the-art DNN model, the CNN+CSP model \cite{cai2020low}, achieved an accuracy of 82.1\% with a 5 second decision window, whereas our result improved upon the accuracy by 12\%, used a 5x smaller decision window, and 8x fewer EEG electrodes. Hence, the overall gains in using lesser number of electrodes, reducing the memory footprint by using fewer parameters with lower bit precision and reducing processing latency by using a shorter decision window, advocate the use of corticomorphic architectures when processing biological signals such as EEG, to detect auditory attention. 

\begin{table}[!h]
    \centering
    \begin{adjustbox}{width=0.45\columnwidth, center}\begin{tabular}{|c|c|c|c|c|}
        \hline\xrowht{15pt}
         Model & \pbox{10cm}{Weight Bit\\ Precision} & \pbox{10cm}{Number of\\ Parameters} & Accuracy & F1-Score\\
         \hline\xrowht{8pt}
         CNN-SNN & 32 & 32,442 & 91.03\% & 91.07\%\\
         \hline\xrowht{8pt}
         CNN & 32 & 27,362 & 79.01\% & 79.04\%\\
         \hline\xrowht{8pt}
         CNN-SNN & 16 & 32,442 & 91.77\% & 91.78\% \\
         \hline\xrowht{8pt}
         CNN-SNN & 32 & 23,132 & 88.41\% & 88.70\% \\
         \hline\xrowht{8pt}
         CNN-SNN & 16 & 23,132 & 89.65\% & 89.28\% \\
         \hline
    \end{tabular}
    \end{adjustbox}
    \vspace{2mm}
    \caption{Comparison of the accuracy and F1 scores for the reference CNN model and CNN-SNN model with variations in bit precision for the weights and number of parameters in the model. All experiments in this table consider the ideal case, using 8 EEG channels with a 1 second decision window, averaged across 20 random initialization of training, testing and validation runs.}
    \label{tab:BitandParam}
\end{table}

With the eventual goal of a hardware implementation of the CNN-SNN model for the use of AAD hearing aids, additional experiments were conducted to highlight the ability of the model to require less power and a smaller memory footprint than an ANN model. The initial results shown in Table \ref{tab:8channels} and Table \ref{tab:16channels} uses a 32 bit precision for the weights and 32,363 parameters; these results use the same bit precision as in the CNN model and require more parameters compared to the 27,422 used for the CNN model. This is not ideal for a hardware implementation as it would require the same memory footprint due to the same bit precision and would use more power due to more memory fetches from the larger number of parameters. In order to demonstrate the proposed architecture's ability to be ported on hardware, the bit precision of the weights and the number of parameters in the model were scaled for the case of 8 EEG channels with a 1 second decision window. The results of this experiment are tabulated in Table \ref{tab:BitandParam}.

The initial results for the CNN-SNN and CNN models are in the first and second row, respectively, of the aforementioned table. From these two results it can be seen that the CNN-SNN requires approximately 5,000 more parameters than the CNN model, while using the same bit precision. While the CNN-SNN model does achieve a higher accuracy, as previously mentioned the CNN model would be better suited for hardware in terms of power usage and memory footprints as the original results stands. In order to reduce the memory footprints for a hardware design the bit precision of the weights needs to be reduced. The CNN-SNN model was trained as described in the CNN-SNN section but following training the model's weight precision was reduced to 16 bits and run on the test set. From Table \ref{tab:BitandParam} the results show that when reducing the bit precision in half the accuracy of the model is similar to that of the initial result with no statistical significance in the difference of the accuracies. Furthermore, the power usage of a model can be reduced by decreasing the parameters of the model as this requires less memory fetches. The number of output channels from the CNN layer were scaled from 40 to 30, which in turn reduces the SNN layer size (Input:60, Hidden:60, Output:2)  as well. Training the down-scaled CNN-SNN model with 23,123 parameters, which is less than the CNN model, and 32 bit precision for the weights the model was again able to achieve a similar accuracy to the base model with no statistical significance in the differences. Combining these results, the CNN-SNN model was trained using 16 bit precision and 23,132 parameters. This model is able to achieve a slight decrease in accuracy compared to the initial result but, use 50\% as many bits for the weights and approximately 15.5\% less parameters while achieving an accuracy that is around 10\% higher than the CNN model.

By reducing the number of parameters and bit precision of the CNN-SNN model there is minor trade off in accuracy but, the CNN-SNN model is now viable for a hardware implementation. The memory footprint will be decreased due to the 16 bit precision and the power usage will be lower because of the 4,000 parameter reduction along with the sparse computation of the SNN layers when compared to the CNN model. Not only is the CNN-SNN model able to achieve state of the art results in accuracy it is also the only DNN model that can feasible deployed into a BECI system, such as AAD hearing aids.



\section{Conclusions}
\label{summary}
This paper proposes a corticomorphic architecture inspired by the layered cortical organization of the auditory cortex, to successfully implement low-latency low-footprint auditory attention detection using EEG and speech data from multiple speakers. Implementing real-time AAD capable of deployment on low-power hardware for edge computing is challenging. Hardware implementation on-the-edge has severe power and resource constraints due to smaller form factors, while real-time operation demands fast processing of signals. Other state-of-the-art models lack in one or more criterion and hence are not capable of hardware realization. This paper takes a significant step towards achieving this target by developing, implementing and evaluating a low-footprint, low-latency hybrid AI architecture inspired by the auditory cortex.

The proposed model architecture has surpassed the current state-of-the-art results in modeling the auditory attention network for AAD. The CNN-SNN architecture performs within a more desirable decision window, uses significantly fewer EEG channels, and increases the accuracy compared to current statistical and DNN models. These results indicate that with the proposed CNN-SNN architecture, it is possible to accurately decode the attention of a listener with high accuracy and simultaneously utilize less power consumption and smaller memory footprints as compared to a conventional ANN. Furthermore, by selecting fewer EEG channels with closer proximity to the auditory cortex, it is possible to decode the attention with similar accuracy, since the difference is statistically insignificant. The CNN-SNN has decreasing accuracy as the time window for inference increased. This decrease in accuracy is not a concern as a smaller decision window is desirable for a stand alone in-ear hearing aid for AAD. Hence all functional and performance metrics reinforce our hypothesis that the corticomorphic CNN-SNN architecture is capable of auditory attention decoding in a reasonable decision window while significantly reducing memory usage and possibly overall power consumption.

Furthermore, additional improvements can be made to more accurately model the signal processing pipeline of the top-down attention network in the brain. By replacing the fully-connected spiking layer with a spiking recurrent neural network (SRNN), such as a spiking Long Short-Term Memory (SLSTM), the classification accuracy might be improved further. The addition of an SRNN allows for memory to be formed from each time step which will make the architecture more bio-realistic in modeling the cortical organization of the auditory attention network. Additionally, changing to a fully spiking architecture such as a spiking convolutional-recurrent network (SCRNN) \cite{xing2020new} the model architecture would be entirely bio-realistic making it ultra-low powered, an important characteristic for BECI devices. However, training a SRNN or SCRNN would be challenging and may not result in high classification accuracy using the existing training algorithms for spiking networks.

The overarching goal of this work is to realize a physical system for realtime AAD in hearing aids. Towards that goal, the hybrid CNN-SNN model can be improved further to make it more adept at AAD. The CNN-SNN model, currently implemented  in software using Nengo DL library, is readily portable to energy-efficient reconfigurable hardware such as Field Programmable Gate Array (FPGA) or Intel's Loihi \cite{davies2018loihi} platform. Our immediate next step is to translate the proposed architecture to one of these platforms. In order to fully understand the power consumption and memory footprint benefits, the ultimate goal is to design a custom ASIC with the most optimized fully spiking architecture, to achieve the best performance in all metrics including classification accuracy, network size and weight precision, processing latency, memory footprint and compute resources. The hardware realization of a corticomorphic architecture used in a smart hearing aid will provide great impetus to further study the possibility of brain-inspired processing for bio-realistic modeling in many other applications.

\bibliographystyle{IEEEtran}
\bibliography{sources}

\begin{thebibliography}{10}
\providecommand{\url}[1]{#1}
\csname url@samestyle\endcsname
\providecommand{\newblock}{\relax}
\providecommand{\bibinfo}[2]{#2}
\providecommand{\BIBentrySTDinterwordspacing}{\spaceskip=0pt\relax}
\providecommand{\BIBentryALTinterwordstretchfactor}{4}
\providecommand{\BIBentryALTinterwordspacing}{\spaceskip=\fontdimen2\font plus
\BIBentryALTinterwordstretchfactor\fontdimen3\font minus
  \fontdimen4\font\relax}
\providecommand{\BIBforeignlanguage}[2]{{%
\expandafter\ifx\csname l@#1\endcsname\relax
\typeout{** WARNING: IEEEtran.bst: No hyphenation pattern has been}%
\typeout{** loaded for the language `#1'. Using the pattern for}%
\typeout{** the default language instead.}%
\else
\language=\csname l@#1\endcsname
\fi
#2}}
\providecommand{\BIBdecl}{\relax}
\BIBdecl

\bibitem{kochkin2009marketrak}
S.~Kochkin, ``Marketrak viii: 25-year trends in the hearing health market,''
  \emph{Hearing review}, vol.~16, no.~11, pp. 12--31, 2009.

\bibitem{cherry1953some}
E.~C. Cherry, ``Some experiments on the recognition of speech, with one and
  with two ears,'' \emph{The Journal of the acoustical society of America},
  vol.~25, no.~5, pp. 975--979, 1953.

\bibitem{shinn2008selective}
B.~G. Shinn-Cunningham and V.~Best, ``Selective attention in normal and
  impaired hearing,'' \emph{Trends in amplification}, vol.~12, no.~4, pp.
  283--299, 2008.

\bibitem{taylor2015does}
B.~Taylor and D.~Hayes, ``Does current hearing aid technology meet the needs of
  healthy aging,'' \emph{Hearing Review}, vol.~22, no.~2, pp. 22--26, 2015.

\bibitem{kong2014classification}
Y.-Y. Kong, A.~Mullangi, and K.~Kokkinakis, ``Classification of fricative
  consonants for speech enhancement in hearing devices,'' \emph{PloS one},
  vol.~9, no.~4, p. e95001, 2014.

\bibitem{Dil01}
H.~Dillon, \emph{Hearing aids}.\hskip 1em plus 0.5em minus 0.4em\relax
  Boomerang Press, 2001.

\bibitem{Ric00}
T.~Ricketts, ``Impact of noise source configuration on directional hearing aid
  benefit and performance,'' \emph{Ear and Hearing}, vol.~21, no.~3, pp.
  194--205, 2000.

\bibitem{Ham02}
T.~A. Powers and V.~Hamacher, ``Three-microphone instrument is designed to
  extend benefits of directionality,'' \emph{The Hearing Journal}, vol.~55,
  no.~10, pp. 38--40, 2002.

\bibitem{Doe96}
M.~Dorbecker and S.~Ernst, ``Combination of two--channel spectral subtraction
  and adaptive wiener post--filtering for noise reduction and
  dereverberation,'' in \emph{in European Signal Processing Conference
  (EUSIPCO}, 1996.

\bibitem{Kol93}
B.~Kollmeier, J.~Peissig, and V.~Hohmann, ``Binaural noise-reduction hearing
  aid scheme with real-time processing in the frequency domain.''
  \emph{Scandinavian Audiology. Supplementum}, vol.~38, pp. 28--38, 1992.

\bibitem{Mar94}
R.~Martin, ``Cspectral subtraction based on minimum statistics,'' in \emph{in
  European Signal Processing Conference (EUSIPCO}, 1994.

\bibitem{Ber79}
M.~Berouti, R.~Schwartz, and J.~Makhoul, ``Enhancement of speech corrupted by
  acoustic noise,'' in \emph{IEEE International Conference on Acoustics,
  Speech, and Signal Processing, ICASSP}, vol.~4, 1979, pp. 208--211.

\bibitem{Eph84}
Y.~Ephraim and D.~Malah, ``Speech enhancement using a minimum-mean square error
  short-time spectral amplitude estimator,'' \emph{IEEE Transactions on
  Acoustics, Speech and Signal Processing,}, vol.~32, no.~6, pp. 1109--1121,
  1984.

\bibitem{Mar02}
R.~Martin, ``Speech enhancement using mmse short time spectral estimation with
  gamma distributed speech priors,'' in \emph{IEEE International Conference on
  Acoustics, Speech, and Signal Processing (ICASSP)}, 2002.

\bibitem{Mar03n}
R.~Martin and C.~Breithaupt, ``Speech enhancement in the dft domain using
  laplacian speech priors,'' in \emph{in Proc. International Workshop on
  Acoustic Echo and Noise Control (IWAENC 03}, 2003.

\bibitem{Lot03}
T.~Lotter and P.~Vary, ``Noise reduction by maximum a posteriori spectral
  amplitude estimation with supergaussian speech modeling,'' in \emph{in Proc.
  International Workshop on Acoustic Echo and Noise Control (IWAENC’03},
  2003.

\bibitem{Sch97}
T.~Schneider and R.~Brennan, ``A multichannel compression strategy for a
  digital hearing aid,'' in \emph{IEEE International Conference on Acoustics,
  Speech, and Signal Processing, ICASSP}, 1997, pp. 411--414.

\bibitem{Cor95}
L.~E. Cornelisse, R.~C. Seewald, and D.~G. Jamieson, ``The input/output
  formula: A theoretical approach to the fitting of personal amplification
  devices,'' \emph{The Journal of the Acoustical Society of America}, vol.~97,
  no.~3, pp. 1854--1864, 1995.

\bibitem{Byr01}
D.~Byrne, H.~Dillon, T.~Ching, R.~Katsch, and G.~Keidser, ``Nal-nl1 procedure
  for fitting nonlinear hearing aids: Characteristics and comparisons with
  other procedures,'' \emph{Journal of the American Academy of Audiology},
  vol.~12, no.~1, pp. 37--51, 2001.

\bibitem{Mar01}
P.~J. Blamey, L.~Martin, H.~Fiket, K.~Krauze, E.~Saunders, B.~Steele, M.~Los,
  B.~Dickson, and P.~Jay, ``Adaptive dynamic range optimization for hearing
  aids,'' in \emph{9th Westem Pacific Acoustics Conference}, 2006, pp. 26--28.

\bibitem{kong2012development}
Y.-Y. Kong and A.~Mullangi, ``On the development of a frequency-lowering system
  that enhances place-of-articulation perception,'' \emph{Speech
  communication}, vol.~54, no.~1, pp. 147--160, 2012.

\bibitem{kong2013using}
------, ``Using a vocoder-based frequency-lowering method and spectral
  enhancement to improve place-of-articulation perception for hearing-impaired
  listeners,'' \emph{Ear and hearing}, vol.~34, no.~3, p. 300, 2013.

\bibitem{Kat97}
J.~M. Kates, ``Classification of background noises for hearing-aid
  applications,'' \emph{Journal of the Acoustical Society of America}, vol.~97,
  no.~1, pp. 461--470, 1995.

\bibitem{Pel02}
V.~Peltonen, J.~Tuomi, A.~Klapuri, J.~Huopaniemi, and T.~Sorsa, ``Computational
  auditory scene recognition,'' in \emph{IEEE International Conference on
  Acoustics, Speech, and Signal Processing (ICASSP)}, vol.~2, 2002, pp.
  II--1941.

\bibitem{Bre90}
A.~S. Bregman, \emph{Auditory scene analysis: The perceptual organization of
  sound}.\hskip 1em plus 0.5em minus 0.4em\relax MIT press, 1994.

\bibitem{Wu03}
M.~Wu, D.~Wang, and G.~J. Brown, ``A multipitch tracking algorithm for noisy
  speech,'' \emph{IEEE Transactions on Speech and Audio Processing}, vol.~11,
  no.~3, pp. 229--241, 2003.

\bibitem{kong2015effects}
Y.-Y. Kong, G.~Donaldson, and A.~Somarowthu, ``Effects of contextual cues on
  speech recognition in simulated electric-acoustic stimulation,'' \emph{The
  Journal of the Acoustical Society of America}, vol. 137, no.~5, pp.
  2846--2857, 2015.

\bibitem{Bij10}
C.~Van~Bijleveld, R.~C. Hendriks, R.~Heusdens, and C.~H. Taal, ``Signal
  processing for hearing aids,'' \emph{Maxwell: Periodiek der Electrotechnische
  Vereeniging, 14 (1)}, 2010.

\bibitem{NIH}
``{NIH Fact Sheet:} {H}earing {A}ids.''

\bibitem{Woo53}
N.~L. Wood and N.~Cowan, ``The cocktail party phenomenon revisited: attention
  and memory in the classic selective listening procedure of cherry (1953).''
  \emph{Journal of Experimental Psychology: General}, vol. 124, no.~3, p. 243,
  1995.

\bibitem{Ham04}
V.~Hamacher, J.~Chalupper, J.~Eggers, E.~Fischer, U.~Kornagel, H.~Puder, and
  U.~Rass, ``Signal processing in high-end hearing aids: State of the art,
  challenges, and future trends,'' \emph{EURASIP Journal on Applied Signal
  Processing}, vol. 2005, pp. 2915--2929, 2005.

\bibitem{biesmans2015comparison}
W.~Biesmans, J.~Vanthornhout, J.~Wouters, M.~Moonen, T.~Francart, and
  A.~Bertrand, ``Comparison of speech envelope extraction methods for eeg-based
  auditory attention detection in a cocktail party scenario,'' in \emph{2015
  37th annual international conference of the ieee engineering in medicine and
  biology society (embc)}.\hskip 1em plus 0.5em minus 0.4em\relax IEEE, 2015,
  pp. 5155--5158.

\bibitem{biesmans2017auditory}
W.~Biesmans, N.~Das, T.~Francart, and A.~Bertrand, ``Auditory-inspired speech
  envelope extraction methods for improved eeg-based auditory attention
  detection in a cocktail party scenario,'' \emph{IEEE Transactions on Neural
  Systems and Rehabilitation Engineering}, vol.~25, no.~2, pp. 1--11, 2017.

\bibitem{das2016effect}
N.~Das, W.~Biesmans, A.~Bertrand, and T.~Francart, ``The effect of head-related
  filtering and ear-specific decoding bias on auditory attention detection,''
  \emph{Journal of neural engineering}, vol.~13, no.~5, p. 056014, 2016.

\bibitem{yost1997cocktail}
W.~A. Yost, ``The cocktail party problem: Forty years later,'' \emph{Binaural
  and spatial hearing in real and virtual environments}, pp. 329--347, 1997.

\bibitem{yost1993auditory}
W.~A. Yost and S.~Sheft, ``Auditory perception,'' in \emph{Human
  psychophysics}.\hskip 1em plus 0.5em minus 0.4em\relax Springer, 1993, pp.
  193--236.

\bibitem{bronkhorst2015cocktail}
A.~W. Bronkhorst, ``The cocktail-party problem revisited: early processing and
  selection of multi-talker speech,'' \emph{Attention, Perception, \&
  Psychophysics}, vol.~77, no.~5, pp. 1465--1487, 2015.

\bibitem{baluch2011mechanisms}
F.~Baluch and L.~Itti, ``Mechanisms of top-down attention,'' \emph{Trends in
  neurosciences}, vol.~34, no.~4, pp. 210--224, 2011.

\bibitem{horton2014envelope}
C.~Horton, R.~Srinivasan, and M.~D’Zmura, ``Envelope responses in
  single-trial eeg indicate attended speaker in a ‘cocktail party’,''
  \emph{Journal of neural engineering}, vol.~11, no.~4, p. 046015, 2014.

\bibitem{o2015attentional}
J.~A. O'sullivan, A.~J. Power, N.~Mesgarani, S.~Rajaram, J.~J. Foxe, B.~G.
  Shinn-Cunningham, M.~Slaney, S.~A. Shamma, and E.~C. Lalor, ``Attentional
  selection in a cocktail party environment can be decoded from single-trial
  eeg,'' \emph{Cerebral cortex}, vol.~25, no.~7, pp. 1697--1706, 2015.

\bibitem{di2015low}
G.~M. Di~Liberto, J.~A. O’Sullivan, and E.~C. Lalor, ``Low-frequency cortical
  entrainment to speech reflects phoneme-level processing,'' \emph{Current
  Biology}, vol.~25, no.~19, pp. 2457--2465, 2015.

\bibitem{ding2014robust}
N.~Ding, M.~Chatterjee, and J.~Z. Simon, ``Robust cortical entrainment to the
  speech envelope relies on the spectro-temporal fine structure,''
  \emph{Neuroimage}, vol.~88, pp. 41--46, 2014.

\bibitem{ding2012emergence}
N.~Ding and J.~Z. Simon, ``Emergence of neural encoding of auditory objects
  while listening to competing speakers,'' \emph{Proceedings of the National
  Academy of Sciences}, vol. 109, no.~29, pp. 11\,854--11\,859, 2012.

\bibitem{kong2014differential}
Y.-Y. Kong, A.~Mullangi, and N.~Ding, ``Differential modulation of auditory
  responses to attended and unattended speech in different listening
  conditions,'' \emph{Hearing research}, vol. 316, pp. 73--81, 2014.

\bibitem{aiken2008human}
S.~J. Aiken and T.~W. Picton, ``Human cortical responses to the speech
  envelope,'' \emph{Ear and hearing}, vol.~29, no.~2, pp. 139--157, 2008.

\bibitem{lalor2010neural}
E.~C. Lalor and J.~J. Foxe, ``Neural responses to uninterrupted natural speech
  can be extracted with precise temporal resolution,'' \emph{European journal
  of neuroscience}, vol.~31, no.~1, pp. 189--193, 2010.

\bibitem{fu2021auditory}
Z.~Fu, B.~Wang, X.~Wu, and J.~Chen, ``Auditory attention decoding from eeg
  using convolutional recurrent neural network,'' in \emph{2021 29th European
  Signal Processing Conference (EUSIPCO)}.\hskip 1em plus 0.5em minus
  0.4em\relax IEEE, 2021, pp. 970--974.

\bibitem{haghighi2017graphical}
M.~Haghighi, M.~Moghadamfalahi, M.~Akcakaya, B.~G. Shinn-Cunningham, and
  D.~Erdogmus, ``A graphical model for online auditory scene modulation using
  eeg evidence for attention,'' \emph{IEEE Transactions on Neural Systems and
  Rehabilitation Engineering}, vol.~25, no.~11, pp. 1970--1977, 2017.

\bibitem{cai2020low}
S.~Cai, E.~Su, Y.~Song, L.~Xie, and H.~Li, ``Low latency auditory attention
  detection with common spatial pattern analysis of eeg signals.'' in
  \emph{INTERSPEECH}, 2020, pp. 2772--2776.

\bibitem{an2021decoding}
W.~W. An, A.~Pei, A.~L. Noyce, and B.~Shinn-Cunningham, ``Decoding auditory
  attention from eeg using a convolutional neural network,'' in \emph{2021 43rd
  Annual International Conference of the IEEE Engineering in Medicine \&
  Biology Society (EMBC)}.\hskip 1em plus 0.5em minus 0.4em\relax IEEE, 2021,
  pp. 6586--6589.

\bibitem{haghighi2018eeg}
M.~Haghighi, M.~Moghadamfalahi, M.~Akcakaya, and D.~Erdogmus, ``Eeg-assisted
  modulation of sound sources in the auditory scene,'' \emph{Biomedical signal
  processing and control}, vol.~39, pp. 263--270, 2018.

\bibitem{o2014power}
J.~A. O’sullivan and J.~Alan, ``Power, nima mesgarani, siddharth rajaram,
  john j foxe, barbara g shinn-cunningham, malcolm slaney, shihab a shamma, and
  edmund c lalor. attentional selection in a cocktail party environment can be
  decoded from single-trial eeg,'' \emph{Cerebral cortex}, vol.~25, no.~7, pp.
  1697--1706, 2014.

\bibitem{mirkovic2015decoding}
B.~Mirkovic, S.~Debener, M.~Jaeger, and M.~De~Vos, ``Decoding the attended
  speech stream with multi-channel eeg: implications for online, daily-life
  applications,'' \emph{Journal of neural engineering}, vol.~12, no.~4, p.
  046007, 2015.

\bibitem{canziani2016analysis}
A.~Canziani, A.~Paszke, and E.~Culurciello, ``An analysis of deep neural
  network models for practical applications,'' \emph{arXiv preprint
  arXiv:1605.07678}, 2016.

\bibitem{xu2018scaling}
X.~Xu, Y.~Ding, S.~X. Hu, M.~Niemier, J.~Cong, Y.~Hu, and Y.~Shi, ``Scaling for
  edge inference of deep neural networks,'' \emph{Nature Electronics}, vol.~1,
  no.~4, pp. 216--222, 2018.

\bibitem{antelis2020spiking}
J.~M. Antelis, L.~E. Falc{\'o}n \emph{et~al.}, ``Spiking neural networks
  applied to the classification of motor tasks in eeg signals,'' \emph{Neural
  networks}, vol. 122, pp. 130--143, 2020.

\bibitem{qiao2015reconfigurable}
N.~Qiao, H.~Mostafa, F.~Corradi, M.~Osswald, F.~Stefanini, D.~Sumislawska, and
  G.~Indiveri, ``A reconfigurable on-line learning spiking neuromorphic
  processor comprising 256 neurons and 128k synapses,'' \emph{Frontiers in
  neuroscience}, vol.~9, p. 141, 2015.

\bibitem{essera2016convolutional}
S.~K. Essera, P.~A. Merollaa, J.~V. Arthura, A.~S. Cassidya, R.~Appuswamya,
  A.~Andreopoulosa, D.~J. Berga, J.~L. McKinstrya, T.~Melanoa, D.~R. Barcha
  \emph{et~al.}, ``Convolutional networks for fast energy-efficient
  neuromorphic computing,'' \emph{Proc. Nat. Acad. Sci. USA}, vol. 113, no.~41,
  pp. 11\,441--11\,446, 2016.

\bibitem{zhang2020neuro}
W.~Zhang, B.~Gao, J.~Tang, P.~Yao, S.~Yu, M.-F. Chang, H.-J. Yoo, H.~Qian, and
  H.~Wu, ``Neuro-inspired computing chips,'' \emph{Nature electronics}, vol.~3,
  no.~7, pp. 371--382, 2020.

\bibitem{KELL2018630}
\BIBentryALTinterwordspacing
A.~J. Kell, D.~L. Yamins, E.~N. Shook, S.~V. Norman-Haignere, and J.~H.
  McDermott, ``A task-optimized neural network replicates human auditory
  behavior, predicts brain responses, and reveals a cortical processing
  hierarchy,'' \emph{Neuron}, vol.~98, no.~3, pp. 630--644.e16, 2018. [Online].
  Available:
  \url{https://www.sciencedirect.com/science/article/pii/S0896627318302502}
\BIBentrySTDinterwordspacing

\bibitem{drakopoulos2021convolutional}
F.~Drakopoulos, D.~Baby, and S.~Verhulst, ``A convolutional neural-network
  framework for modelling auditory sensory cells and synapses,''
  \emph{Communications Biology}, vol.~4, no.~1, pp. 1--17, 2021.

\bibitem{pickles1998introduction}
J.~Pickles, ``An introduction to the physiology of hearing,'' in \emph{An
  Introduction to the Physiology of Hearing}.\hskip 1em plus 0.5em minus
  0.4em\relax Brill, 1998.

\bibitem{chechik2012auditory}
G.~Chechik and I.~Nelken, ``Auditory abstraction from spectro-temporal features
  to coding auditory entities,'' \emph{Proceedings of the National Academy of
  Sciences}, vol. 109, no.~46, pp. 18\,968--18\,973, 2012.

\bibitem{hunsberger2016training}
E.~Hunsberger and C.~Eliasmith, ``Training spiking deep networks for
  neuromorphic hardware,'' \emph{arXiv preprint arXiv:1611.05141}, 2016.

\bibitem{Rasmussen2018}
\BIBentryALTinterwordspacing
D.~Rasmussen, ``{NengoDL}: Combining deep learning and neuromorphic modelling
  methods,'' \emph{arXiv}, vol. 1805.11144, pp. 1--22, 2018. [Online].
  Available: \url{http://arxiv.org/abs/1805.11144}
\BIBentrySTDinterwordspacing

\bibitem{xing2020new}
Y.~Xing, G.~Di~Caterina, and J.~Soraghan, ``A new spiking convolutional
  recurrent neural network (scrnn) with applications to event-based hand
  gesture recognition,'' \emph{Frontiers in neuroscience}, vol.~14, p. 1143,
  2020.

\bibitem{davies2018loihi}
M.~Davies, N.~Srinivasa, T.-H. Lin, G.~Chinya, Y.~Cao, S.~H. Choday, G.~Dimou,
  P.~Joshi, N.~Imam, S.~Jain \emph{et~al.}, ``Loihi: A neuromorphic manycore
  processor with on-chip learning,'' \emph{Ieee Micro}, vol.~38, no.~1, pp.
  82--99, 2018.

\end{thebibliography}

\end{document}